# The $T_c$ amplification by quantum interference effects in diborides

Annette Bussmann-Holder

*Max-Planck Institut f r Festk rperforschung*
*Heisenbergerstrasse 1 D 70759 Stuttgart Germany*

Antonio Bianconi

*Dipartimento di Fisica and Istituto Nazionale per la Fisica della Materia,*
*Universit di Roma "La Sapienza", P. le Aldo Moro 2, 00185 Roma, Italy*

## Abstract

The model of two ( $\sigma$ and $\pi$) channel superconductivity known to be necessary to explain the superconductivity in $MgB_2$ has been applied to the $Al_{1-x}Mg_xB_2$ diborides by tuning x from $MgB_2$ to $AlMgB_4$. The evolution of the interband coupling parameter (probing the strength of the interchannel pairing due to quantum interference effects) and the two gaps in the $\sigma$ and $\pi$ channel as a function of x have been calculated. While in $MgB_2$ the quantum interference effects give an amplification of $T_c$ by a factor of 1.5 in comparison with the dominant intra $\sigma$ band single channel pairing, in $AlMgB_4$ the amplification is about 100, in comparison with the dominant intra $\pi$ band single channel pairing.

For correspondence email addresses:
bianconi@superstripes.com
A.Bussmann-Holder@fkf.mpg.de

The interest in enhancing the superconducting transition temperature $T_c$ by quantum interference between the pairing in two channels is receiving renewed attention even though its theory has been established in the fifties [1] and extended later by several authors for conventional [2,3] and non conventional superconductors [4-6] including multi channel interactions. In fact the recent experimental results for $MgB_2$ show that the two band model is needed to explain both the normal and the superconducting properties [7-19]. Here the key ingredient that allows the physical realization of the process of two channel $T_c$ amplification is the fact that the two gaps refer to two different parts in k space (a large gap in the σ Fermi surface and a small gap in the π Fermi surface) and are well separated in real space, one for the σ holes in the boron layers, and the other for π electrons in the interstitial space (Mg layers between the boron layers). Therefore the system can be described as a multi layer structure made of alternating layers of metallic and superconducting planes. While in most of the materials (like in conventional isotropic 3D superconductors) the impurity interband scattering suppresses the quantum interference, here it is very low and does not suppress the two band superconductivity enhancement [19].

It has been found that the $Al_{1-x}Mg_xB_2$ alloys [20-31] show a continuos evolution through a complicated mixed phase from $MgB_2$ (x=1) to the end member $AlMgB_4$ (x=0.5) where a well ordered superlattice structure of boron layers intercalated by alternating layers of Al and Mg is formed [20-24]. In spite of the fact that the alloys with intermediate x are rather disordered, the superconducting transition temperature is still well defined and it drops systematically with decreasing x as shown in Fig. 1a [from refs. 25,29,30 in agreement with several independent experiments refs. 21-24]. The x-dependence of $T_c$ shows a kink around x=0.7 where a dimensionality crossover of the Fermi surface takes place [31,25-26] and $T_c$ reaches 3K at x=0.5 for $AlMgB_4$. At x=0.7 also the partial density of states of the σ band (Fig.1b) shows a kink. The superconducting phase in the ordered phase $AlMgB_4$ is a highly interesting case since the Fermi level is driven near the top of the σ band, the partial density of states (PDOS) in the σ band is strongly reduced and the Fermi energy $E_F$ for the σ holes is only about 100-200 meV [25-27].

The two-gap scenario has already been invoked for pure $MgB_2$ and electron-phonon interactions together with Coulomb potentials are well known [14, 16, 17]. The interesting question here is whether such an approach can be used for the alloys up to the ordered $AlMgB_4$ phase and how the gap structure and interband couplings develop with doping.

Going from x=1 to x=0.5 a dramatic increase in the $E_{2g}$ phonon mode energy $\omega_{E2g}$ takes place [29-31] increasing from 70 meV to 115 meV as shown in Fig. 2a (from the data in ref. 30) indicating a strong reduction in the electron-phonon interaction which reflects itself also in a reduction of the phonon damping as shown in Fig. 2b (from the data in ref. 30) while the average phonon frequencies remain nearly constant $\omega_{ln}$ =59-62 meV. The $E_{2g}$ phonon mode is known to couple strongly with the modulation of the top of the σ band and defining the electron-phonon coupling [33,37] like:

$$\lambda_\sigma = 2N_\sigma(E_f)\left[\frac{h}{2M_B\omega_{E_{2g}}{}^2}\right]\left|\sum_{j=1,2}\hat{\varepsilon}_j \cdot \vec{D}_j\right|^2 \qquad (1)$$

where D=130 meV/pm is the deformation potential [33] and $N_\sigma(E_f)$ is the partial density of states at the Fermi level in the σ band, in MgB$_2$ reaches the strong or intermediate coupling regime $\lambda_\sigma \approx 1$ [33-36]. On the other hand the electron phonon coupling for the π electrons remains in the weak coupling limit: $\lambda_\pi = 0.44$. While in the early days of the research on superconductivity in MgB$_2$ it was suggested that the large value of $\lambda_\sigma$ was sufficient to explain the high T$_c$ within the standard single band isotropic Migdal-Eliasberg approach, is has been recognized recently that this approach fails and that it is necessary to consider a two band model [16]. By using equation (1) to obtain $\lambda_\sigma(x)$ and $\lambda_\pi(x)$ and considering Coulomb pseudopotentials $\mu_\sigma(x)$ and $\mu_\pi(x)$ that are normalized at x=1 to the values given in ref.16 we have calculated T$_c$ within the McMillan or Allen-Dynes approach. The calculated superconducting transition temperatures for two ideal different metals made of only σ and π electrons respectively, without interband interactions, are shown in Fig. 3. The screened effective couplings including the Coulomb shielding $\lambda_1(x)$ and $\lambda_2(x)$, respectively, which are the inverse of the exponent in the McMillan equation, are plotted in Fig. 4. From Fig. 3 we can clearly see that the single band model fails to predict the experimental T$_c$(x) dependence in fact it is obvious that a single band approach fails to explain the T$_c$ of 3K observed in AlMgB$_4$ and is also far beyond the experimental data for varying x. This motivated us to model also the alloys within the two-band scenario within the framework of the BCS approximation using the experimental data for the two intraband couplings $\lambda_1(x)$ and $\lambda_2(x)$ as shown in Fig. 4.

Starting from the values of the energies of the phonon modes $\omega_{E2g}(x)$ and $\omega_{ln}(x)$, the experimental values of T$_c$(x), $\lambda_1(x)$ and $\lambda_2(x)$, we have obtained the interband coupling $\lambda_{12}(x)$ and the intra band gaps $\Delta_1(x)$ and $\Delta_2(x)$ for the σ and π band, respectively. The Hamiltonian we use, reads:

$$H = H_0 + H_1 + H_2 + H_{12} \tag{2}$$

$$H_0 = \sum_{k_1\sigma} \xi_{k_1} \sigma^+_{k_1\sigma} \sigma_{k_1\sigma} + \sum_{k_2\sigma} \xi_{k_2} \pi^+_{k_2\sigma} \pi_{k_2\sigma} \tag{2a}$$

$$H_1 = -\sum_{k_1 k_1' q} V_1(k_1,k_1') \sigma^+_{k_1+q/2\uparrow} \sigma^+_{-k_1+q/2\downarrow} \sigma_{-k_1'+q/2\downarrow} \sigma_{k_1'+q/2\uparrow} \tag{2b}$$

$$H_2 = -\sum_{k_2 k_2' q} V_2(k_2,k_2') \pi^+_{k_2+q/2\uparrow} \pi^+_{-k_2+q/2\downarrow} \pi_{-k_2'+q/2\downarrow} \pi_{k_2'+q/2\uparrow} \tag{2c}$$

$$H_{12} = -\sum_{k_1 k_2 q} V_{12}(k_1,k_2) \{\sigma^+_{k_1+q/2\uparrow} \sigma^+_{-k_1+q/2\downarrow} \pi_{-k_2+q/2\downarrow} \pi_{k_2+q/2\uparrow} + h.c.\}, \tag{2d}$$

where $H_0$ is the kinetic energy of the bands $i=1,2$ with $\xi_{k_i} = \varepsilon_i + \varepsilon_{k_i} - \mu$. Here $\varepsilon_i$ denotes the position of the σ-band and π-band band with creation and annihilation operators $\sigma^+, \sigma, \pi^+, \pi$, respectively, and $\mu$ is the chemical potential. The pairing potentials $V_i(k_i,k_i')$ act intraband and $V_{12}(k_1,k_2)$ is the interband interaction which is dominated by multiphonon processes. By performing a BCS mean field analysis of Equs. 2 and applying standard techniques, we obtain:

$$< \sigma_{k_1-}^+ \sigma_{-k_1}^+ J \; > = \frac{\overline{\Delta}_{k_1}}{2E_{k_1}} \tanh[\frac{\beta E_{k_1}}{2}] = \overline{\Delta}_{k_1} \Phi_{k_1} \tag{3a}$$

$$< \pi_{k_2-}^+ \pi_{-k_2}^+ J \; > = \frac{\overline{\Delta}_{k_2}}{2E_{k_2}} \tanh[\frac{\beta E_{k_2}}{2}] = \overline{\Delta}_{k_2} \Phi_{k_2} \tag{3b}$$

with $E_{k_1}^2 = \xi_{k_1}^2 + |\overline{\Delta}_{k_1}|^2, \overline{\Delta}_{k_1} = \Delta_{k_1} + A_{k_1}$ and $E_{k_2}^2 = \xi_{k_2}^2 + |\overline{\Delta}_{k_2}|^2, \overline{\Delta}_{k_2} = \Delta_{k_2} + B_{k_2}$. From this the selfconsistent set of equations for the coupled gaps is given by:

$$\overline{\Delta}_{k_1} = \sum_{k_1'} V_1(k_1,k_1')\overline{\Delta}_{k_1'} \Phi_{k_1'} + \sum_{k_2} V_{1,2}(k_1,k_2)\overline{\Delta}_{k_2} \Phi_{k_2} \tag{4a}$$

$$\overline{\Delta}_{k_2} = \sum_{k_2'} V_2(k_2,k_2')\overline{\Delta}_{k_2'} \Phi_{k_2'} + \sum_{k_1} V_{1,2}(k_1,k_2)\overline{\Delta}_{k_1} \Phi_{k_1} \tag{4b}$$

which have to be solved simultaneously for each temperature and gap value. Starting from experimental values of the doping dependence of the phonon frequencies together with the effective electron-phonon interactions $\lambda_{ii} = N_{ii}(0)V_i(k_i,k_i')$ (i=1,2) for the intraband pairing processes, the values of the interband couplings are adjusted to fit the experimental values of $T_c$ as already outlined above. The results are shown in Fig. 4 where the effective electron-phonon couplings are presented. The corresponding energy gaps at T=0 K are shown in Fig.5a and the gap to $T_c$ ratios are depicted in Fig. 5b. As it is well known for the two-band model, both gap to $T_c$ ratios deviate substantially from BCS predictions – one being strongly enhanced, while the other is far below the values predicted by BCS theory. The calculated value of $\lambda_{12}$(x=1) is consistent with the average value of the screened coupling constants $\lambda_{\sigma\pi}$ and $\lambda_{\pi\sigma}$ derived form the two band model of MgB$_2$ [16] using the corresponding values of the pseudopotentials $\mu_{\sigma\pi}$ and $\mu_{\pi\sigma}$. The obtained values of $\Delta_1$(x=1) and $\Delta_2$(x=1) are in very good agreement with the superconducting gaps measured by Ivarrone et al. for MgB$_2$ [8].

We observe that the interband coupling parameter $\lambda_{12}$(x), as shown in Fig. 4 increases by decreasing x from 1 (MgB$_2$), to reach a maximum for values of x near 0.6-0.7 where the strength of the interchannel pairing due to quantum interference effects is maximum. In the range x= 0.6-0.7 the $T_c$(x) curve shows a kink which is the signature that the Fermi level has been tuned at the cross-over of the Fermi surface of the $\sigma$-band from 2D to 3D dimensionality. This is the expected position of the "shape resonance" [33]. The two gaps in the $\sigma$ and $\pi$ channel as a function of x plotted in Fig.5 show a very interesting case of interchange of their dominance and a gap crossing takes place at x=0.6 where the $\sigma$-band related gap becomes smaller than the $\pi$ related gap. For AlMgB$_4$ we have therefore a different physical situation for the two gap scenario. In fact in MgB$_2$ the interchannel interference effects push $T_c$ up to the strong coupling regime (2$\Delta_1$/$T_c$=4.2) with an effective amplification of $T_c$ of the order of 1.5-2 increasing the strong-intermediate coupling regime of the dominant 2D $\sigma$ band. In AlMgB$_2$ it is the 3D $\pi$ band with the dominant gap being 2 dimensional which is supported by the 3D $\sigma$ band with a smaller intraband coupling constant $_{-1}$. While the intraband pairing processes alone will give $T_c$ in the range of 1-10 milliK the actual $T_c$ is 3K, i.e. an amplification of a factor 100-1000 is realised in this scenario.

As a consequence of the interchange of the driving band going through the "shape resonance" at x=0.6-0.7 the gap separation is strongly doping dependent, being very large for $MgB_2$, intermediate for x=0.75 and reversed at x=0.5. The temperature dependence of the gaps for the above mentioned three cases is shown in Fig. 6 where substantial differences are predicted for the three cases which can be tested by further experiments.

In conclusion, we have shown that also the alloyed systems $Al_{1-x}Mg_xB_2$ with 0.5≤x≤1.0 are best described within a two-band model where interband interactions are the dominant force that drives $T_c$ to the experimentally observed values. Predictions for the ratio of the two gaps are made, where especially a reverse in gap magnitudes is obtained.

One of A.B. would like to thank the Max-Planck-Institutes for their hospitality where this work has been accomplished. This work is supported by "progetto cofinanziamento Leghe e composti intermetallici: stabilità termodinamica, proprietà fisiche e reattività" of MIUR, and by "Progetto 5% Superconduttività" of Consiglio Nazionale delle Ricerche (CNR).

.

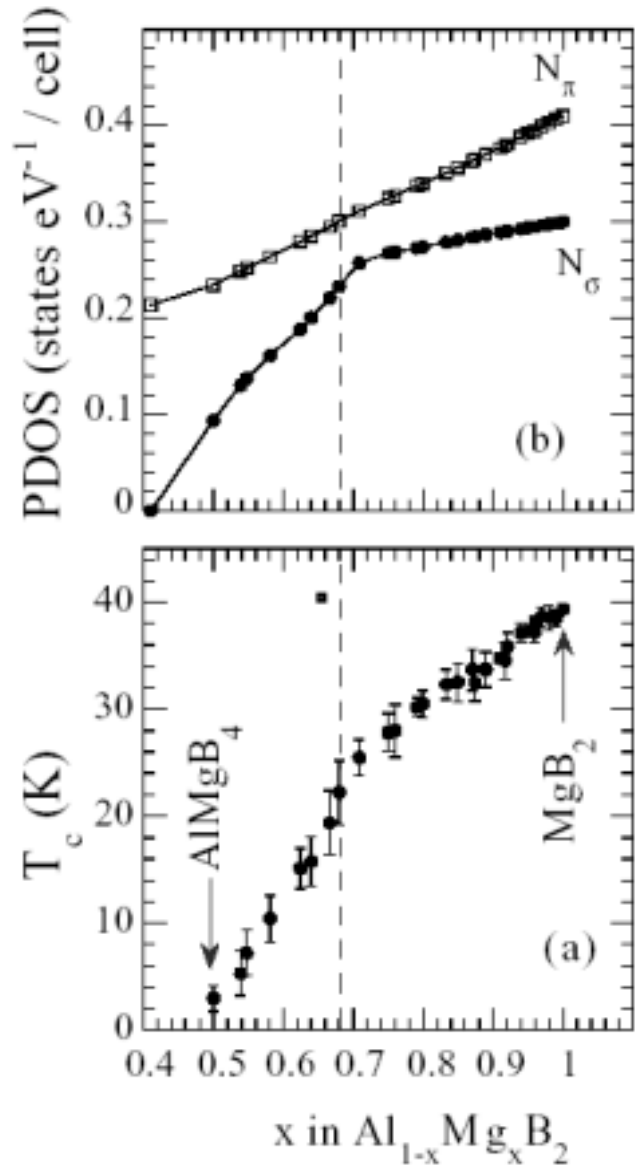

**Fig.1** (panel a) The superconducting transition temperature for $Al_{1-x}Mg_xB_2$ from x=0.5 (AlMgB$_4$) to x=1 (MgB$_2$) from ref. 25,29,30. (panel b) The partial density of states (PDOS) of the $\sigma$ band $N_\sigma$ and of the $\pi$ band $N_\pi$ as function of x.

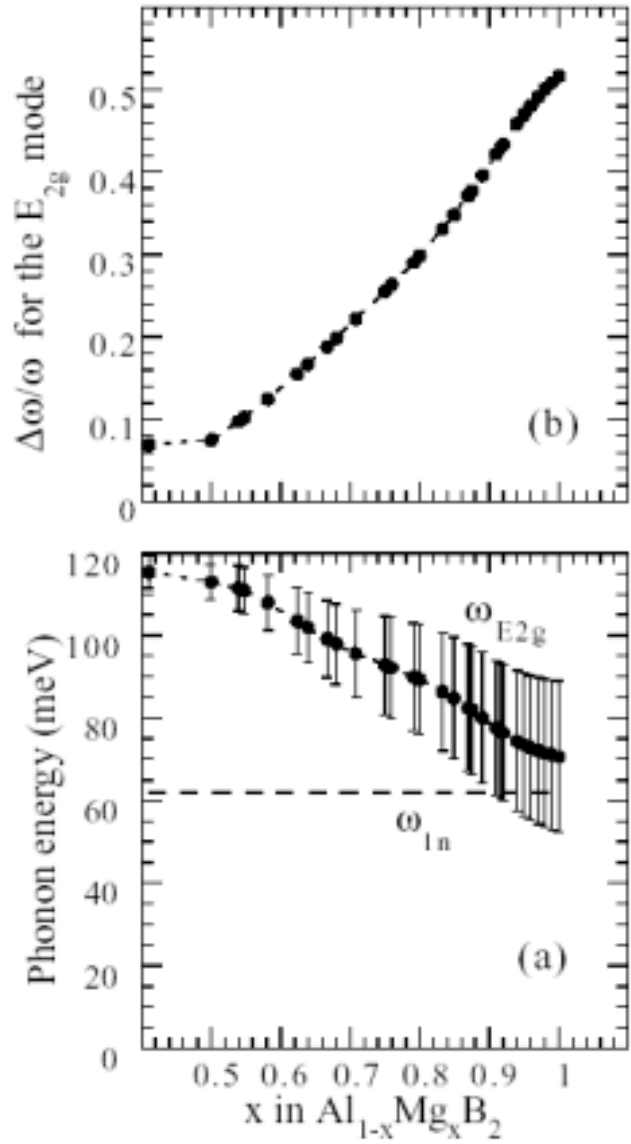

**Fig. 2** (panel a) The variation of the energy $\omega_{E2g}$ of the $E_{2g}$ phonon mode as function of x from x=1 to x=0.5. The average phonon frequencies remain nearly constant $\omega_{ln}$ =62 meV . (panel b) The $E_{2g}$ phonon damping, defined as the ratio of the total width of the average energy of the Raman line as a function of x (from the data in ref. 30).

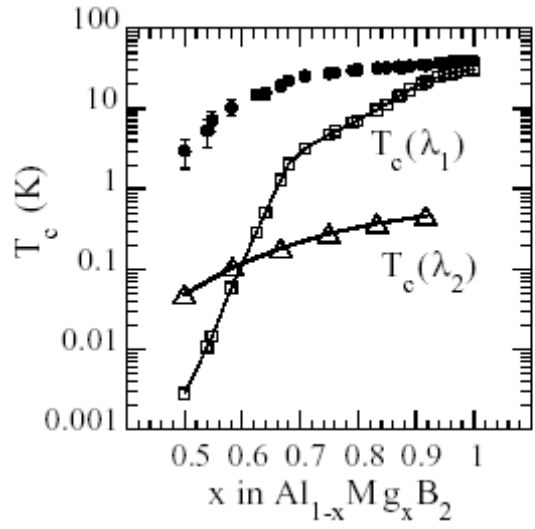

**Fig. 3** The calculated superconducting transition temperature $T_c$ for two ideal systems made of of $\sigma$ and $\pi$ electrons only are compared with the experimental data. The single band isotropic Migdal-Eliasberg approach has been considered and $T_c$ has been calculated using the McMillan or Allen-Dynes formula considering Coulomb pseudo potentials $\mu_\sigma(x)$ and $\mu_\pi(x)$ and electron phonon interactions normalized at x=1 to the values given in ref.16.

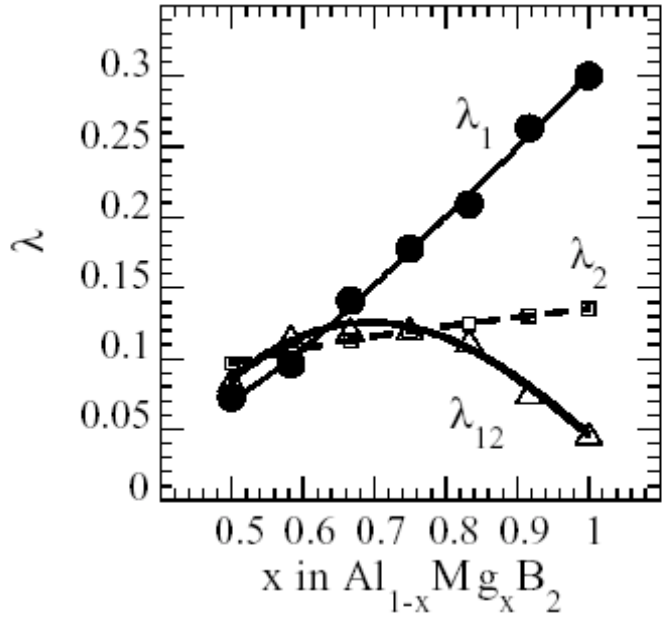

**Fig. 4** The screened effective couplings $\lambda_1(x)$ and $\lambda_2(x)$ for the $\sigma$ and $\pi$ electrons respectively which are the inverse of the exponent in the McMillan equation. The interband coupling $\lambda_{12}$ has been calculated by using the two band interference model in such a way as to reproduce the experimental values of $T_c$ for each value of x.

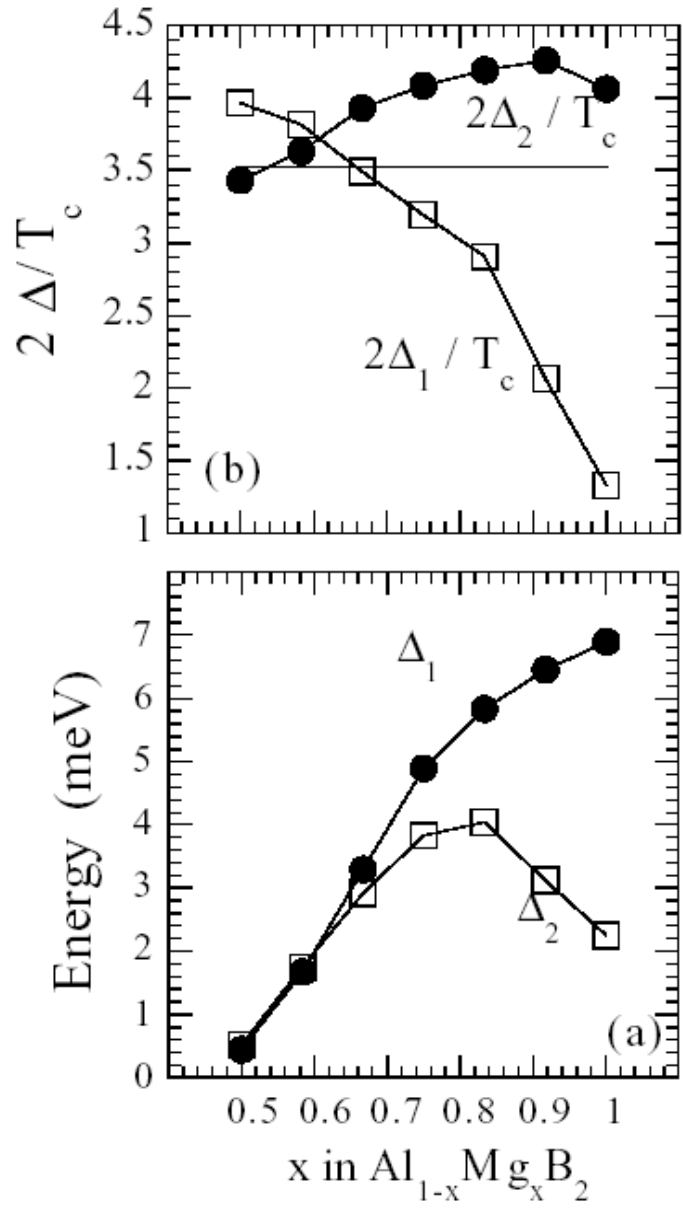

**Fig. 5** (panel a) The energy gaps, at T=0K, $\Delta_1(x)$ and $\Delta_2(x)$ for the $\sigma$ and $\pi$ electrons, respectively, as obtained by solving equations 3 and 4 simultaneously and selfconsistently, and (panel b) the corresponding gap to $T_c$ ratios as a function of x.

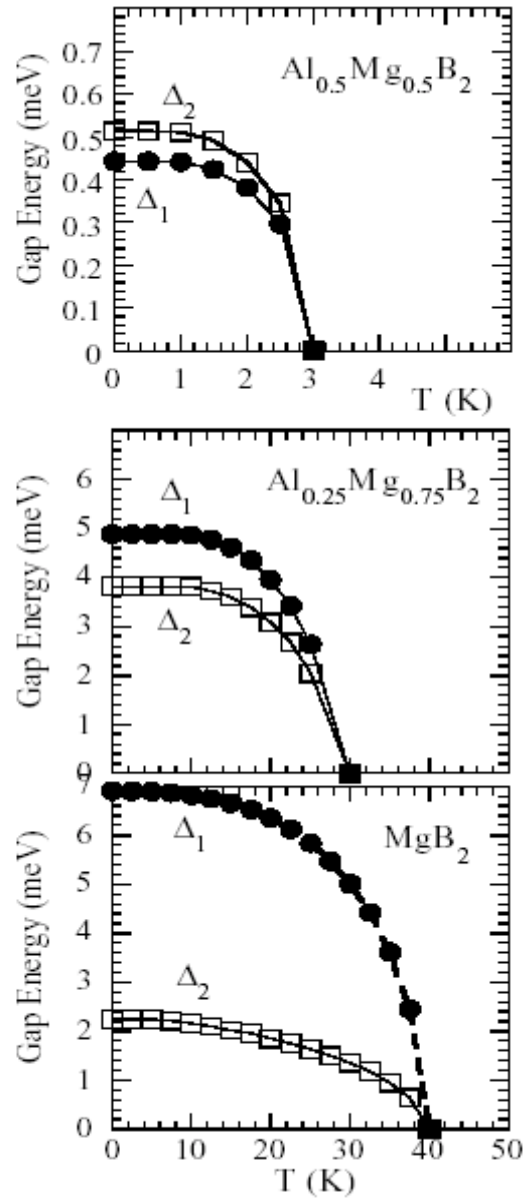

**Fig. 6** The predicted temperature dependence of the gaps for three different systems $MgB_2$ (top panel), $Al_{0.25}Mg_{0.75}B_2$ (middle panel) and $AlMgB_4$ (lower panel).